\documentstyle[12pt]{article}

\newcommand{\bb}{\begin{equation}}
\newcommand{\ee}{\end{equation}}
\newcommand{\m}{\mbox{$\frac{1}{2}$}}
\date{}

\title{The Black Hole in Three Dimensional Spacetime}
\author{ M\'aximo Ba\~nados$^{(1,a)}$ Claudio Teitelboim$^{(1,2,a)}$
and Jorge Zanelli$^{(1,a)}$ \\
\small {\em $~^{(1)}$Centro de Estudios Cient\'{\i}ficos de Santiago,
Casilla 16443, Santiago 9, Chile} \\ \small {\em and Facultad de Ciencias,
Universidad de Chile, Casilla 653, Santiago, Chile} \\ \small {\em $^{(2)}
$Institute for Advanced Study,
Olden Lane, Princeton, New Jersey 08540, USA.} \normalsize}
\begin{document}
\maketitle

\begin{abstract}
The standard Einstein-Maxwell equations in 2+1 spacetime
dimensions, with a negative cosmological constant, admit a black
hole solution. The 2+1 black hole -characterized by mass, angular
momentum and charge, defined by flux integrals at infinity- is
quite similar to its 3+1 counterpart.  Anti-de Sitter space
appears as a negative energy state separated by a mass gap from
the continuous black hole spectrum.  Evaluation of the partition
function yields that the entropy is equal to twice the perimeter
length of the horizon.
\end{abstract}

The fascinating properties of the black hole, classical and
-especially- quantum, have made it long desirable to have
available a lower dimensional analog which could exhibit the
key features without the unnecessary complications.

It is the purpose of this letter to report that the sought for
analog does exist in standard 2+1 Einstein-Maxwell theory with
a negative cosmological constant.

For simplicity we will first ignore the coupling to the Maxwell
field. The generalization to non-zero electric charge will be
indicated afterwards.

The action is

\bb
I= \frac{1}{2\pi} \int \sqrt{-g} \left[ R + 2l^{-2}
\right]d^2 xdt  + B,
\label{1}
\ee
where $B$ is a surface term and the radius $l$ is related to
the cosmological constant by $-\Lambda = l^{-2}$.

The equations of motion derived from (\ref{1}) are solved by the
black hole field

\bb
ds^2 = -N^2 dt^2 + N^{-2} dr^2  + r^2(N^{\phi} dt + d\phi)^2
\label{2}
\ee
where the squared lapse $N^2(r)$  and  the angular shift $N^{\phi}(r)$ are given
by

\begin{eqnarray}
N^2(r) &=& - M + \frac{r^2}{l^2} + \frac{J^2}{4r^2} \label{2.2} \nonumber \\
N^{\phi}(r) &=& - \frac{J}{2r^2}   \nonumber
\end{eqnarray}
with $-\infty < t <\infty$,  $0 < r < \infty$  and  $0 \leq \phi \leq 2\pi$.

In this letter we will focus our attention mainly on the
physical properties of the solution. The geometric structure
will be briefly touched upon at the end and its detailed study
will be provided in a forthcoming publication\cite{bhg}.

The two constants of integration $M$ and $J$ appearing
in (\ref{2}) are the conserved charges
associated with asymptotic invariance under time displacements
(mass) and rotational invariance (angular momentum),
respectively. These charges are given by flux integrals through
a large circle at spacelike infinity.

The lapse function $N(r)$ vanishes for two values of $r$ given by

$$
r_{\pm}=l \left[ \frac{M}{2} \left( 1 \pm \sqrt{1-\left(
\frac{J}{Ml}\right)^2} \right) \right]^{1/2}.
$$
Of these, $r_{+}$ is the black hole horizon. In order for the horizon to
exist one must have

\bb
    M>0,  \;\;\;\;  |J| \leq Ml.
\label{7}
\ee
In the extreme case $|J|=Ml$, both roots of $N^2=0$ coincide.

Note that the radius of curvature $l=(-\Lambda)^{-1/2}$ provides the
length scale necessary in order to have a horizon in a theory in which the
mass is dimensionless. If one lets $l$ grow very large the black hole
exterior is pushed away to infinity and one is left just with the inside.

The vacuum state, namely what is to be regarded as empty space, is
obtained by making the black hole disappear. That is, by letting
the horizon size go to zero. This amounts to letting
$M\rightarrow 0$, which requires $J\rightarrow 0$ on account of
(\ref{7}). One thus obtains the line element

\bb
ds^2_{vac} = -(r/l)^2 dt^2 + (r/l)^{-2} dr^2 + r^2 d\phi^2.
\label{8}
\ee

As $M$ grows negative one encounters the solutions studied
previously in Refs. \cite{1,2}. The conical singularity that
they posses is naked, just as the curvature singularity of a
negative mass black hole in $3+1$ dimensions.  Thus, they must,
in the present context, be excluded from the physical spectrum.
There is however an important exceptional case. When one reaches
$M=-1$ and $J=0$ the singularity dissapears.  There is no
horizon, but there is no singularity to hide either.  The
configuration

$$ ds^2=-(1+(r/l)^2) dt^2 + (1+(r/l)^2)^{-1} dr^2 + r^2 d\phi^2 $$
(anti-de Sitter space) is again permissible.

Therefore, one sees that anti-de Sitter space emerges as a ``bound state",
separated from the continuous black hole spectrum by a mass gap of one
unit. This state cannot be deformed continuously into the vacuum
(\ref{8}), because the deformation would require going through a sequence
of naked singularities which are not included in the configuration space.

Note that the zero point of energy has been set so that the mass vanishes
when the horizon size goes to zero. This is quite natural. It is what is
done in 3+1 dimensions. In the past, the zero of energy has been adjusted
so that, instead, anti-de Sitter space has zero mass. Quite appart from
this diference, the key point is that the black hole spectrum lies above
the limiting case $M=0$.

The $2+1$ black hole has thermodynamic properties similar to those found
in $3+1$ dimensions\cite{4}.  In the steepest descent approximation, the
free energy $F$ divided by the temperature is given by the value of the
Euclidean action evaluated on the Euclidean continuation\cite{f4} of the
black hole field (\ref{2}). The surface terms appearing in the action are
here crucial. They must be constructed so that the action truly has an
extremum on the class of fields considered\cite{3}.  In the variation one
must allow changes in the fields contributing to the surface integrals
giving $M$ and $J$, but must hold fixed their momenta (appropiate
variational derivatives of the action on the boundaries), which become the
``thermodynamical conjugates"\cite{JY}. These conjugates are the period $\beta$ of
the Euclidean Killing time (inverse temperature $T^{-1}$) and the
rotational chemical potential -which turns out to be the negative of the
angular shift $N^{\phi}$ evaluated on the horizon (``angular velocity").

To determine the surface terms, we found it best, both for conceptual and
practical reasons, not to work with the covariant form of the action
(\ref{1}) but to start instead with its Hamiltonian version

$$
I'= \int\left[ \pi^{ij}\dot{g}_{ij} - N{\cal H} -N^i{\cal
H}_i \right] d^2xdt + B'.
$$

The surface term $B'$ differs from $B$ in
(\ref{1}) (the volume integrals of $I$ and $I'$ differ by a
surface term).

Working with the Hamiltonian action has the following
advantages: (i) Since the metric is time independent, the value
of the volume piece of the Hamiltonian action is equal to zero
when the constraints hold.  Thus, the surface terms are
everything, even in the presence of the cosmological constant.
(ii) One knows right away the surface term that must be added at
infinity without need to regularize. For the Euclidean action,
it is simply the period $\beta$ of
Killing time multiplied by the mass (by definition of the
mass).

After infinity has been dealt with, there remains only to make
sure that the variational derivative of the action should vanish
on the horizon. This makes it necessary to include in $B'_{Euc}$ two
further ``surface terms" at $r=r_{+}$.  They turn out to be
equal to minus two times the proper perimeter length of the
horizon (to cancel the variation of the hamiltonian constraint)
and $ \beta N^{\phi}(r_+)  J$ (to cancel the
variation of the momentum constraint).

One thus gets for the Euclidean action

\bb
I_{Euc} = \beta M - 4\pi r_{+} + \beta N^{\phi}(r_+) J.
\label{10}
\ee
   But, $I_{Euc} = F/T$, where the free energy is $F=M-TS-\sum \mu_i
C_i$ and the $\mu$'s are the chemical potentials
thermodynamically conjugate to the conserved charges $C_i$.
Therefore, equation (\ref{10}) confirms that $\beta$ and
$-N^{\phi}(r_{+})$ are the inverse temperature and the chemical potential
corresponding to $J$, respectively. It also shows
that the entropy is equal to twice the perimeter length of the
horizon,

\bb
S= 2 L = 4\pi r_{+}.
\label{12}
\ee

From (\ref{12}), one may evaluate the temperature of the black
hole,

$$
T = \left[ \frac{\partial S}{\partial M} \right]_{J}^{-1} =
\frac{r_{+}^2-r_{-}^2}{2\pi r_{+}}.
$$

This expression coincides with the periodicity in Euclidean Killing time
needed to make the Euclidean black hole geometry regular at the horizon.
One may also verify that $N^{\phi}(r_{+})=T (\partial S/
\partial J)_{M}$.

Note that as the horizon disappears, the temperature goes to zero in
contrast with the $3+1$ case. On the other hand, the extreme rotating hole
($J=Ml$) has zero temperature and non-zero entropy, just as the 3+1 case.

Now, we briefly discuss how the electromagnetic field is brought in. One
includes the following additional contributions in the action: (i) The
electromagnetic energy and momentum densities are added to ${\cal H}$ and
${\cal H}_i$ respectively, (ii) A term $\pi^i \dot{A}_i$ is added to
$\pi^{ij}\dot{g}_{ij}$, (iii) The Gauss law constraint is incorporated by
adding $+ \int d^2x dt A_0 \pi^i_{,i}$ to the volume piece of the action.
(iv) This makes it mandatory to include in $B_{Euc}'$ a new surface
integral equal to $A_0(r_{+}) Q$.  Here $Q$ is the electric charge given
by a flux integral at infinity, and equal to the constant value throughout
space of the radial component $\pi^r$.

The only non vanishing component of the electromagnetic vector potential
may be taken to be

$$
A_0 (r) = - Q ln(r/r_0).
$$

The only modification of the metric (\ref{2}) is that the lapse
function in (\ref{2.2}) must be replaced by

$$
N^2 = N^{2}_{(Q=0)} + \m Q A_0(r).
$$
The free energy acquires an extra term $-A_0(r_+) Q$ and the entropy
is again equal to twice the proper perimeter length of the
black hole.  The horizon exists for any value of $Q$ provided the
bound (\ref{7}) on $J$ is obeyed\cite{f6}.

Lastly, we turn to some comments on the geometry of the black hole.  For
simplicity, these comments are restricted to $Q=0$ (no Maxwell field).  In
that case, one is dealing with a spacetime of constant negative curvature
(the Riemann tensor is a constant multiple of an antisymmetrized product
of metric tensors).  It is well known \cite{witten}  that such a space
time must arise from identifications of points in anti de Sitter space
through a discrete subgroup of its symmetry group $O(2,2)$.  In this case,
the discrete subgroup is generated by one element, the exponential of a
particular Killing vector. In terms of the embedding

$$ -u^2 - v^2 + x^2 + y^2 = -l^2 $$
of anti de Sitter space in flat four dimensional space that
Killing vector is given by

$$\xi = \frac{r_{+}}{l} \left( x\frac{\partial}{\partial u}
+ u\frac{\partial}{\partial x} \right) - \frac{r_{-}}{l} \left(
y\frac{\partial}{\partial v}  + v \frac{\partial}{\partial y} \right).$$

Throughout anti de Sitter space this vector can be spacelike, null or
timelike. The whole of the black hole geometry is the region where $\xi$
is spacelike. This region is incomplete. Its boundaries are the surfaces
$\xi^2=0$ which correspond to $r=0$ in the metric (\ref{2}). One cannot
continue past these boundaries because $\xi$ becomes timelike and the
identification would produce closed timelike lines.

The rich structure of the $2+1$ black hole is remarkable given
the simple nature of gravitation in three spacetime dimensions\cite{9}.
One may hope that its study will provide further understanding
of the black hole, especially at the quantum level.

Discussions with David Brown, Marc Henneaux, Steven Carlip,
Frank Wilczek, Edward Witten and James York are gratefully acknowledged. M.
B. holds a Fundaci\'on Andes Fellowship and C. T. holds a
Guggenheim Fellowship.  This work was supported in part by
grants 0862/91 and 0867/91 of FONDECYT, (Chile), by a European
Communities research contract, and by institutional support
provided by SAREC (Sweden) and Empresas Copec (Chile) to the
Centro de Estudios Cient\'{\i}ficos de Santiago.

\newpage

\noindent  {\bf Note Added:} The
charged, rotating solution as described in this Letter is incorrect, as
pointed out by several authors [11].  The correct metric can be found in
[12] and had been independently obtained in a different approach in [13].

\begin{center}
\rule{3cm}{.4mm}
\end{center}


\begin{thebibliography}{22}
\bibitem{f4} The Euclidean continuation
             may be obtained by setting $t=-i\tau$ in (\ref{2}). This makes the
             metric
             complex when $J\neq 0$. The Euclidean action is given by
             $iI=-I_{Euc}$. Although correct in the case at hand, making
             imaginary a particular coordinate in a generally covariant
             theory is disturbing. A more satisfactory
             way to define the Euclidean continuation is the
             following. One solves the Euclidean field equations but
             introduces complex constants of integration in the solutions,
             thus generating complex solutions to real equations. In the
             present example, one sets $J_{Euc} = -i J$, and, when there is
             electric charge, $Q_{Euc}=-iQ$. Here $J$ and $Q$ are the
             physical (Minkowskian signature) values of the angular momentum
             and electric charge. This point of view makes it clear that the
             procedure is coordinate invariant.
\bibitem{f6} The constant $r_0$ is an arbitrarly reference point for the
             energy, which cannot be taken to be at
	     infinity as in the $3+1$ case. If $r_0$ is replaced by $r_0'$, $M$ is
	     replaced
	     by $M + \m Q^2 ln(r_0' / r_0)$. However, the variation of
	     the free energy keeping the temperature and the chemical
	     potentials fixed is independent of $r_0$, and so are,
	     $T$, $S$ and $N^{\phi}(r_+)$.
\end{thebibliography}
\end{document}